# Thermal conductivity of molybdenum disulfide nanotube from molecular dynamics simulations

Han Meng[1], Dengke Ma[1,2], Xiaoxiang Yu[1], Lifa Zhang[2], Zhijia Sun[3], Nuo Yang[1,] *

[1] State Key Laboratory of Coal Combustion, School of Energy and Power Engineering, Huazhong University of Science and Technology, Wuhan 430074, China

[2] NNU-SULI Thermal Energy Research Center (NSTER) & Center for Quantum Transport and Thermal Energy Science (CQTES), School of Physics and Technology, Nanjing Normal University, Nanjing 210023, China

[3] Institute of High Energy Physics, Chinese Academy of Sciences, Beijing 100049, China

* Corresponding E-mail: nuo@hust.edu.cn (N. Y.)

# Abstract

Single layer molybdenum disulfide (SLMoS$_2$), a semiconductor possesses intrinsic bandgap and high electron mobility, has attracted great attention due to its unique electronic, optical, mechanical and thermal properties. Although thermal conductivity of SLMoS$_2$ has been widely investigated recently, less studies focus on molybdenum disulfide nanotube (MoS$_2$NT). Here, the comprehensive temperature, size and strain effect on thermal conductivity of MoS$_2$NT are investigated. A chirality-dependent strain effect is identified in thermal conductivity of zigzag nanotube, in which the phonon group velocity can be significantly reduced by strain. Besides, results show that thermal conductivity has a ~$T^{-1}$ and a ~$L^{\beta}$ relation with temperature from 200 to 400 K and length from 10 to 320 nm, respectively. This work not only provides feasible strategies to modulate the thermal conductivity of MoS$_2$NT, but also offers useful insights into the fundamental mechanisms that govern the thermal conductivity, which can be used for the thermal management of low dimensional materials in optical, electronic and thermoelectrical devices.

# Introduction

In recent years, two-dimensional single layer molybdenum disulfide (SLMoS$_2$) with a tri-layer structure composed of one layer of Mo atoms sandwiched between two layers of S atoms, has attracted great attention due to its unique electronic[1], optical[2], and mechanical[3, 4] properties. Different from graphene, which is a gapless conductor, SLMoS$_2$ is a semiconductor with an intrinsic bandgap and high electron mobility, which make it become a promising candidate for many electronic and optoelectronic applications[5-7]. Similar to carbon nanotube (CNT), the quasi one-dimensional single walled molybdenum disulfide nanotube (MoS$_2$NT) can be formed by rolling up SLMoS$_2$. On the other hand, thermal properties of low-dimensional systems are very important for not only the performance and reliability of devices, but also the fundamental understanding of the physics.

Thermal properties of SLMoS$_2$ have been widely studied recently. Jin *et al*. calculated the thermal conductivity of SLMoS$_2$ using equilibrium molecular dynamics and reported a value of 116.8 Wm$^{-1}$K$^{-1}$ at room temperature[8]. Wu *et al*. studied the isotropic effect on the thermal conductivity of SLMoS$_2$ and found that Mo isotopes contribute more and can strongly scatter phonons with intermediate frequency for large size samples[9]. Ding *et al*. found that the thermal conductivity of SLMoS$_2$ can be effectively tuned by introducing even a small amount of lattice defects, and can be further tuned by mechanical strain[10]. As for MoS$_2$NT, only a few researches focus on the mechanical properties[11]. The study on thermal conductivity of MoS$_2$NT is still lacking.

In this work, we investigated numerically the thermal conductivities of MoS$_2$NT. Firstly, we constructed two chirality of MoS$_2$NT with different size by rolling up different SLMoS$_2$. Then, we study the temperature, diameter and length effect on thermal conductivity of MoS$_2$NT, as well as the strain effect. Lastly, we perform lattice dynamics analysis to interpret the chirality-dependent strain effect on thermal conductivity by quantifying the phonon dispersion relations and group velocity.

## Model and method

Different type of MoS$_2$NT is constructed by rolling up SLMoS$_2$ based on specific lattice vector **r** = m**a$_1$** + n**a$_2$**, where the lattice constants of the primitive cell are $a_1$= $a_2$=3.147 Å (as shown in figure 1a). Noting that our study only focuses on armchair nanotube (aNT, m=n) and zigzag nanotube (zNT, m≠0 and n=0) without considering chiral MoS$_2$NT. For instance, figure 1b nad 1c show the atomic structure of (8,8) aNT and the (14,0) zNT respectively. To overcome the structural instability induced by small diameter, the (40, 40) aNT and (70, 0) zNT are chosen as the smallest structure in the following calculation.

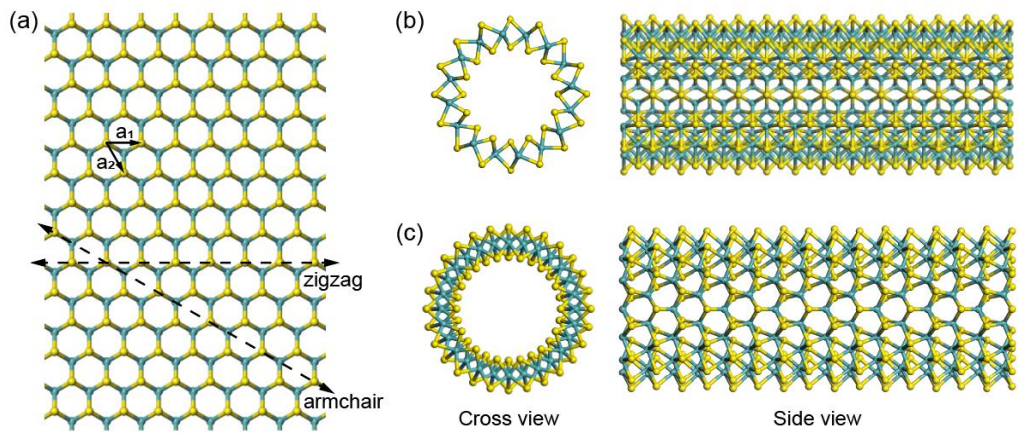

**Figure 1**. Schematic construction of MoS$_2$NT from single layer MoS$_2$. (a) Hexagonal SLMoS$_2$ lattice with indication of primitive vectors and the rolling directions for aNT and zNT. (b-c) The cross and side view of (8,8) aNT and (14,0) zNT. Yellow and green spheres represent S and Mo atoms respectively.

The classical non-equilibrium molecular dynamics (NEMD) method has been employed in the calculation of thermal properties[12-17]. The thermal conductivity is calculated based on the Fourier's law of heat conduction as

$$\kappa = -\frac{J}{A\nabla T},\qquad(1)$$

where $A$ is the cross-sectional area, $\nabla T$ is the temperature gradient, $J$ is the heat flux that recorded by the average of the input and output power of the two baths as

$$J = \frac{\Delta E_{in}+\Delta E_{out}}{2\Delta t},\qquad(2)$$

where $\Delta E$ is the energy added to or removed from each heat bath during each time step $\Delta t$.

All simulations are performed by the large-scale atomic/molecular massively parallel simulator (LAMMPS) package[18]. The interatomic interaction is described by Stillinger-Weber potential, which includes both two-body and three-body terms and has been widely used to study the thermal properties[19, 20]. Time step is set as 0.5 fs, and the velocity Verlet algorithm is used to integrate the discrete differential equations of motion[21]. The fixed and periodic boundary conditions are applied in axial and other two directions, respectively. Two Langevin thermostats with a temperature difference of 20 K are used to establish temperature gradient along axial direction. The cross-section of nanotube is defined as a ring with the thickness of 6.16 Å. To overcome the statistical error, the results are averaged over five independent simulations with different initial conditions. (NEMD simulation details are given in supplementary material)

## Results and discussion

We first calculate the thermal conductivity of $MoS_2NT$ with a diameter of 7 nm and a length of 10 nm at 300 K, and the results are depicted in figure 2a as an illustration of the NEMD method. By linear fitting temperature profile (the red line), temperature gradient is obtained to calculate thermal conductivity. Since temperature usually play a critical role in thermal transport, we further investigate the dependence of thermal conductivity on temperature. As shown in figure 2b, thermal conductivity of aNT is a little higher than that of zNT, with an approximate value of 16 Wm$^{-1}$K$^{-1}$ obtained at room temperature, which is on the same order of magnitude as the results of single layer $MoS_2$[8, 9]and aligned CNT-PE composites[22]. Thermal conductivity of both aNT and zNT decrease when temperature increases from 200 K to 400 K. Moreover, it is obvious that thermal conductivity decreases as $\sim T^{-1}$, which is attributed to the enhancement of Umklapp phonon-phonon scattering at higher temperature. The similar result is also found by previous study on CNT [23].

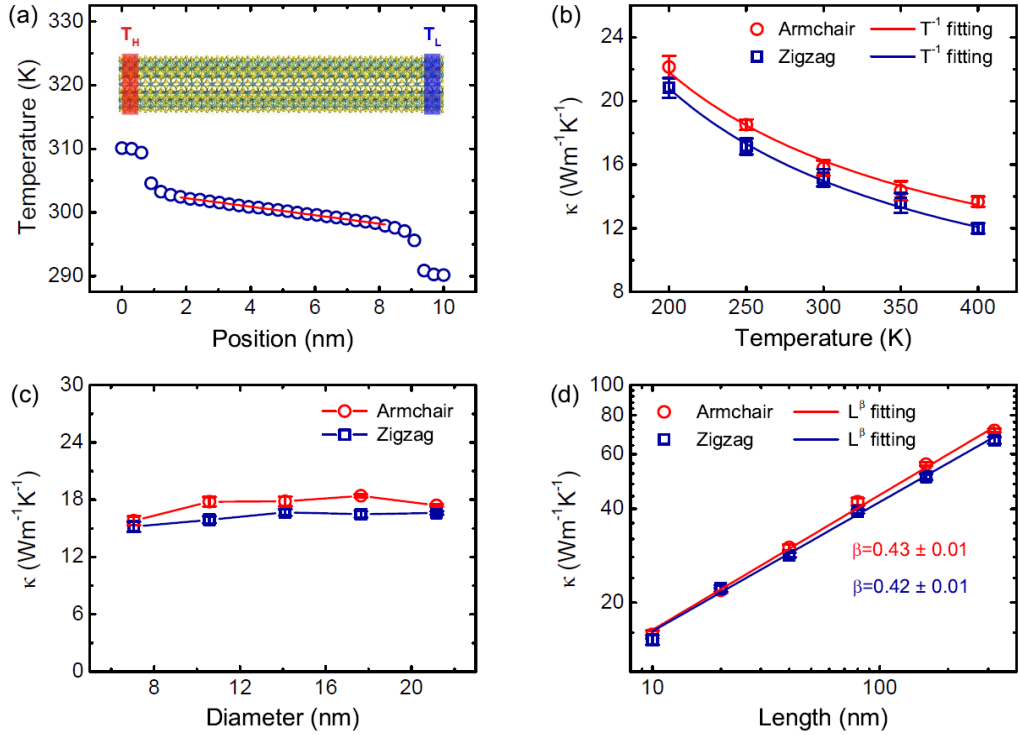

**Figure 2**. (a) The schematic of MoS$_2$NT and linear fitting of the temperature profile obtained from averaging during NEMD simulation; Thermal conductivity of MoS$_2$NT with different chirality versus (b) temperature (the diameter and length are set as 7 nm and 10 nm respectively), (c) length (the diameter is set as 7 nm) and (d) diameter (the length is set as 10 nm) at 300 K.

Then, we study the dependence of thermal conductivity on diameter at room temperature, where the length is fixed as 10 nm. As shown in figure 2c, thermal conductivity of both aNT and zNT almost unchanged with diameter increases. It indicates that thermal conductivity is independent on diameter, which is consistent with the result of CNT[24]. Furthermore, we study the dependence of thermal conductivity on length at room temperature, where the diameter is fixed as 7 nm. As shown in figure 2d, thermal conductivity of both aNT and zNT has a strong dependence on length. With length up to 320 nm, the values are obtained as high as 71 ± 1 Wm$^{-1}$K$^{-1}$ and 66 ± 2 Wm$^{-1}$K$^{-1}$ for aNT and zNT respectively. We fitted the results and it shows that thermal conductivity diverges with length as $\kappa \propto$

$L^\beta$, β is obtained as 0.43 ± 0.01 and 0.42 ± 0.01 for aNT and zNT respectively, which is consistent with the previous value 0.4 obtained from CNT[25]. As studied here, the length of nanotube is longer than the phonon mean free path, the phonon-phonon scattering plays a key role in the process of phonon transport. Therefore, the anomalous heat diffusion induced by diffusive phonon transport is responsible for the length dependent thermal conductivity. The similar length effect was also observed in the previous studies on CNT[23], SLMoS$_2$[9] and silicon nanowires[26].

Besides temperature and size effect, we also investigate the strain effect on thermal conductivity for both aNT and zNT, the diameter and length are fixed as 7 nm and 10 nm respectively. Due to the limitation of structural stability, the strain is applied along the axial direction of nanotube, ranging from -0.03 to 0.03 for aNT and -0.03 to 0.09 for zNT. As shown in figure 3a, aNT can only tolerate strain in a small range and thermal conductivity is insensitive to strain. In contrast, thermal conductivity of zNT is significantly influenced by strain. It decreases almost linearly with tensile strain, but abruptly reduces by about fifty percent when strain is larger than 0.06. That is, the strain effect on thermal conductivity has a strong chirality dependence.

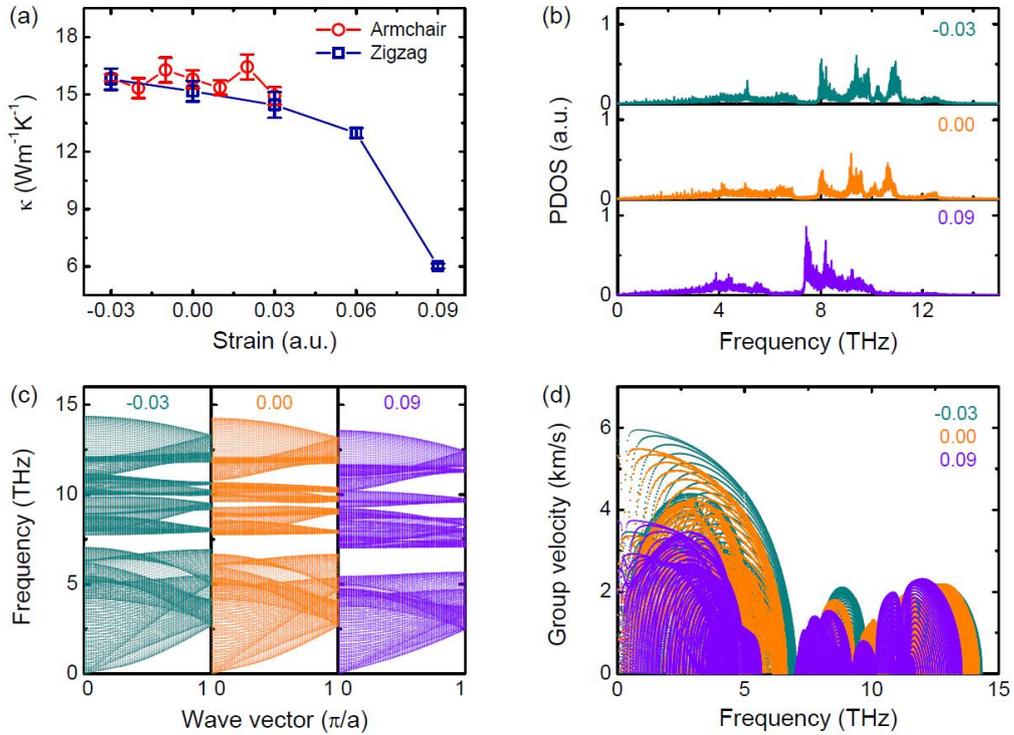

**Figure 3**. (a) Thermal conductivity of MDNT with different chirality versus strain at room temperature, the diameter and length are set as 7 nm and 10 nm respectively; (b) Normalized PDOS along axial direction for different strained zNT. (c) phonon dispersion relation for different strained zNT; (d)

Phonon group velocity versus frequency for different strained zNT.

To understand the underlying mechanism of the strain effect on thermal conductivity, we calculate the phonon density of state (PDOS) along axial direction for different strained nanotube. The PDOS spectra are obtained by Fourier transforming time-domain velocity function. As shown in figure 3b, the lower frequency phonon modes are not sensitive to the strain. Differently, higher frequency modes shift to low-frequency region when nanotube undergoes tensile strain, which indicates that phonon dispersion is compressed to lower frequency region so that phonon group velocity could be reduced. To show it clearly, we calculate the phonon dispersion by the general utility lattice program (GULP)[27]. Figure 3c shows that the phonon modes shift to the lower frequency region, which is consistent with that indicated by PDOS. The phonon group velocity for different strained nanotube are shown in figure 3d. It is obvious that group velocity decreases when nanotube is strained, especially the low frequency part, which mainly contribute to thermal conductivity. On the basis of $\kappa = cv^2\tau/3$, decreasing thermal conductivity is therefore attributed to the reduction of group velocity induced by strain.

## Conclusion

In general, the thermal conductivity of $MoS_2NT$ is systematically studied by NEMD simulations. The results show that thermal conductivity decreases with temperature as $T^{-1}$ from 200 to 400 K, which show the dominant three-phonon scattering mechanism. Moreover, thermal conductivity has a strong dependence on length and the value diverges as $L^\beta$ ($\beta\sim0.4$) when L changes from 10 to 320 nm. However, it has a weak dependence on diameter. More importantly, it is found that the strain can dramatically reduce thermal conductivity of zNT rather than that of aNT. That is, there is a chirality-dependent strain effect. By calculating the PDOS and phonon dispersion relations, the strain effect on thermal conductivity is attributed to the decrease of phonon group velocity induced by strain, which makes phonon modes drift to low-frequency region. The results demonstrate the possibility to modulate the thermal conductivity of $MoS_2NT$ through temperature, length and strain. This work offers useful insights into the fundamental mechanisms that govern the thermal conductivity, which can be used for the thermal management of low dimensional materials in optical, electronic and thermoelectrical devices.

# Conflicts of interest

There are no conflicts of interest to declare.

# Acknowledgements

This work is financially supported by the National Natural Science Foundation of China (No. 51576076 and No. 51711540031), the Natural Science Foundation of Hubei Province (No. 2017CFA046), and the Fundamental Research Funds for the Central Universities, HUST (No. 2019kfyRCPY045). We are grateful to Meng An, Xiao Wan, Wentao Feng and Shichen Deng for useful discussions. The authors thank the National Supercomputing Center in Tianjin (NSCC-TJ) and China Scientific Computing Grid (ScGrid) for providing assistance in computations.

Supplementary material

# Thermal conductivity of molybdenum disulfide nanotube from molecular dynamics simulations

Han Meng[1], Dengke Ma[1, 2], Xiaoxiang Yu[1], Lifa Zhang[2], Zhijia Sun[3], Nuo Yang[1, *]

[1] State Key Laboratory of Coal Combustion, School of Energy and Power Engineering, Huazhong University of Science and Technology, Wuhan 430074, China

[2] NNU-SULI Thermal Energy Research Center (NSTER) & Center for Quantum Transport and Thermal Energy Science (CQTES), School of Physics and Technology, Nanjing Normal University, Nanjing 210023, China

[3] Institute of High Energy Physics, Chinese Academy of Sciences, Beijing 100049, China

* Corresponding E-mail: nuo@hust.edu.cn (N. Y.)

## S1. The NEMD simulation details.

All NEMD simulation details are provided in table S1. The simulation procedure is summarized as follows. After minimization, the NPT ensemble is used to relax the MoS$_2$NT structure for 100 ps. Then, except for the thermostat regions, the rest of the system is simulated by NVE ensemble for 100 ps to reach the steady state. Finally, a longer simulation using NVE ensemble is performed for 5 ns to record and average the temperature and heat flux.

Table S1. NEMD simulation details

| Method | | | Non-Equilibrium molecular dynamics | |
|---|---|---|---|---|
| **Parameters** | | | | |
| **Potential** | Stillinger-Weber | | **Time step** | 0.5 fs |
| **Thermostat** | Langevin | | **Temperature difference** | 20 K |
| **Simulation process** | | | | |
| **Ensemble** | **Setting** | | | **Purpose** |
| NPT | Boundary condition | x, y, z: p, p, p | | Relax structure |
| | Runtime | 100 ps | | |
| NVE | Boundary condition | x, y, z: p, p, f | | Reach steady state |
| | Runtime | 100 ps | | |
| NVE | Boundary condition | x, y, z: p, p, f | | Record information |
| | Runtime | 5 ns | | |
| **Recorded physical quantity** | | | | |
| Temperature | | $\langle E \rangle = \sum_{i=1}^{N} \frac{1}{2} m_i v_i^2 = \frac{3}{2} N k_B T$ | | |
| Heat flux | | $J = \dfrac{\Delta E_{in} + \Delta E_{out}}{2\Delta t}$ | | |
| Thermal conductivity | | $\kappa = -\dfrac{J}{A \nabla T} = -\dfrac{JL}{A \Delta T}$ | | |

## S2. The phonon dispersion relation and group velocity of aNT.

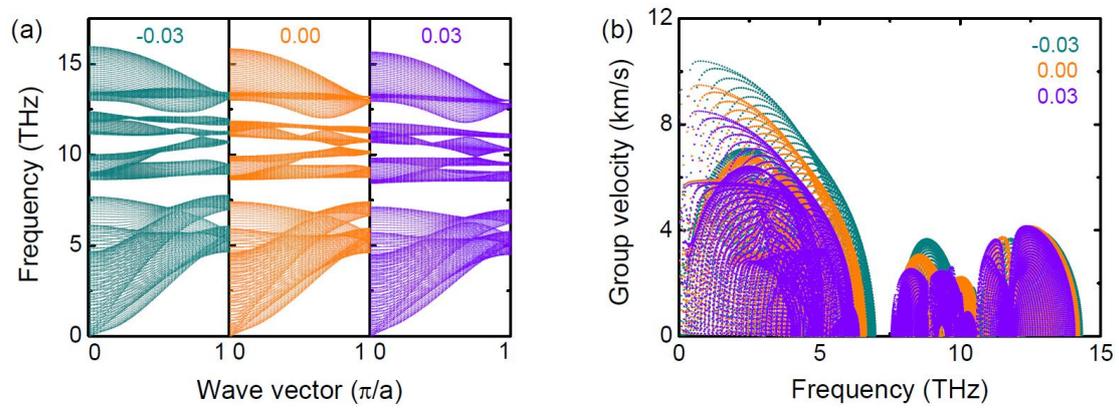

**Figure S1**. (a) phonon dispersion relation for different strained a-NT; (b) Phonon group velocity versus frequency for different strained a-NT.